\documentclass[aps,prl,twocolumn,showpacs,superscriptaddress]{revtex4-1}
\usepackage{amsmath}
\usepackage{epsfig}
\usepackage{amssymb}
\usepackage{dcolumn}
\usepackage{graphicx}
\usepackage{dcolumn}

\usepackage{color}

\newcommand{\beq}{\begin{equation}}
\newcommand{\eeq}{\end{equation}}
\newcommand{\dpsi}{\delta \psi}
\newcommand{\rv}{\mathbf{r}}
\newcommand{\vv}{\mathbf{v}}
\newcommand{\kv}{\mathbf{k}}

\newcommand{\diff}{\mathrm d}

\begin{document}
\date{\today}

\title{Fractional quantum mechanics in polariton condensates with velocity dependent mass}

\author{F. Pinsker}
\affiliation{Department of Physics, Clarendon Laboratory, University of Oxford, Parks Road, Oxford OX1 3PU, United Kingdom}
\email{florian.pinsker@physics.ox.ac.uk}

\author{W. Bao}
\affiliation{Department of Mathematics, National University of Singapore, Singapore 119076}

\author{Y. Zhang}
\affiliation{Beijing Computational Science Research Center, Beijing 100084, P. R. China}
\affiliation{Wolfgang Pauli Institute c/o Fak. Mathematik,
University Wien, Oskar-Morgenstern-Platz 1, 1090 Vienna, Austria}

\author{H. Ohadi}
\affiliation{Department of Physics, Cavendish Laboratory, University of Cambridge, Cambridge CB3 0HE, United Kingdom}

\author{A. Dreismann}
\affiliation{Department of Physics, Cavendish Laboratory, University of Cambridge, Cambridge CB3 0HE, United Kingdom}

\author{J. J. Baumberg}
\affiliation{Department of Physics, Cavendish Laboratory, University of Cambridge, Cambridge CB3 0HE, United Kingdom}

\begin{abstract} We introduce and analyze a novel mean-field model for polariton condensates with  velocity dependence of the effective polariton mass due the photon and exciton components. The effective mass depends on the in-plane wave vector $\kv$, which at the inflection point of the lower polariton energy branch becomes infinite and above negative. The polariton condensate modes of the new mean-field theory are now sensitive to mass variations and for certain points of the energy dispersion the polariton condensate mode represents fractional quantum mechanics. The impact of the generalized kinetic energy term is elucidated by numerical studies in $1$D and $2$D showing significant differences for large velocities.  Analytical expressions for plane wave solutions as well as a linear waves analysis show the significance of this new model.

\end{abstract}

\maketitle

\emph{Introduction.--} About two decades ago the fractional Schr\"odinger equation (FSE) was discovered as a mathematical extension within the Feynman path integral formalism by transposing Brownian with L\`evy-type paths \cite{Luk,Luk2}.  This generalization of the fundamental equation of single-body quantum mechanics has given rise to new intriguing mathematical structures and it forms the base of {\it fractional quantum mechanics} \cite{Luk,Luk2, mark, app, solo, mom}.
The FSE incorporates the concept of an intrinsically nonlocal {\it fractional} kinetic energy,
\begin{eqnarray}\label{Laplace}
(- \Delta)^s f(\rv)&\equiv&\mathcal F^{-1} (|\kv|^{2s} \mathcal F (f)) \nonumber\\
&=&\frac{1}{(2\pi)^d} \int_{\mathbb R^d} |\kv|^{2s} \hat f (\kv) e^{i \kv \cdot \rv} \diff \kv,
\end{eqnarray}
while the linear SE is the special case $s=1$.  $\mathcal F (f) \equiv \hat{f}(\kv) =
\int_{\mathbb R^d} f(\rv)\, e^{-i \kv \cdot \rv}\, \diff \rv$
denotes the Fourier transform of $f(\rv) = \frac{1}{(2\pi)^d}\int_{\mathbb R^d} \hat f(\kv)\,
e^{i \kv \cdot \rv}\, \diff \kv$. On the other hand the concept of velocity dependent mass is well established in solid state physics \cite{Dresselhaus} suggesting a possible route for the implementation of fractional quantum mechanics or even more complex kinetic energies as will be shown in this letter utilizing polariton condensates.

 To introduce the concept of generalized kinetic energy we turn to the solid state system of polariton Bose-Einstein (BE) condensates - macroscopically occupied single mode states that highlight properties of fundamental quantum mechanics ranging from quantum harmonic oscillators \cite{naked, naked2} to interference \cite{interfere, alex} while providing control over key system parameters \cite{opt,Light,Rev1,Pin6}.  We show that the type of kinetic energy in Schr\"odinger like models is of fundamental importance for the modes and particularly for non-equilibrium polariton condensate behavior at different locations of the dispersion. Polariton condensates have kinetic energies of the mathematical form of a Fourier-multiplier, $ \mathcal F^{-1} (g(\kv) \mathcal F (f))$, $g(\kv)$ is a real-valued function associated with the two branches of the polariton dispersion through \cite{Dresselhaus},
\beq\label{bla}
g(\kv) = \frac{|\kv|^2}{m  (\kv)} := g(k)= k^2 \partial_k^2 E_{\rm L,U}(k)
\eeq
with $k=|\kv|$ for $\kv\in {\mathbb R}^d$ and $d=1,2$. The two energy branches of the dispersion curve $E_{\rm L,U}(k)$ vary significantly over $k$ and the kinetic energy \eqref{bla} depends on the $k$ of the injected or spontaneously populated condensate polaritons generally in a non-parabolic way. In fact locally {\it fractional kinetic energies} can be implemented due to the {\it velocity dependent mass} $m  (\kv)$ that modifies the parabolic dispersion accordingly - e.g. a modification of the polariton condensate wave function due to effectively negative mass was recently shown experimentally \cite{Schef}.

In this letter the whole spectrum of the lower polariton branch is considered while taking  the dynamical behaviour into account. We clarify the role of the generalized kinetic energy as it is particularly important for implementations above the inflection point and because several mathematically different forms of the kinetic energy have been used in similar scenarios while neglecting the inherent mathematical inconsistencies of the corresponding predictions as secondary effects \cite{Wout, jona, relax}. Current models catch aspects of the condensate wave function at localized $k$ but the concept introduced here incorporates the mean field treatment for more extended wave packets in $\kv$ space while being the more accurate description even for localized wave packets. Concepts such as energy relaxation can be included in the new PDE \cite{alex, relax}.  Numerically we find that a time splitting Fourier pseudospectral method \cite{Bao1,Bao2} can be used to generate converging solutions, a method that will be presented in more detail in a later work.

\emph{Theoretical Background.--} Polaritons are quasiparticles consisting of excitons and cavity photons within semiconductor micro-cavities which obey Bose-Einstein statistics \cite{Light} and thus the potential to condense into a single particle mode \cite{Kasp}. Excitons are coupled pairs of electrons and holes of oppositely charged spin-half particles in a semiconductor held together by the Coulomb force between them \cite{Rev1}. Excitons interact with light fields \cite{Hopf} and can form integer spin polariton quasiparticles in the strong coupling regime that are confined to the micro-cavity \cite{Weis}.  As polaritons are $10^9$ times lighter than rubidium atoms \cite{Kasp}, condensation is observed in CdTe/CdMgTe/GaAs micro-cavities \cite{Kasp,Light, Ohadi} and recently even at room temperature in flexible polymer-based structures \cite{Plum,Plum2}. The basic Hamiltonian taking the interaction between the cavity light modes and excitons into account is stated in \cite{Rev1, Light}.
By diagonalizing this operator one gets the lower and upper polariton eigenvalues \cite{Rev1},
\begin{multline}\label{energy}
E_{\rm L,U} (\kv) = \frac{1}{2} \bigg(E_{c} (\kv) + E_{x} (\kv) \mp \\ \mp \sqrt{(E_{x} (\kv) - E_{c} (\kv))^2 + 4 \hbar^2 \Omega^2_R} \bigg).
\end{multline}
The dispersion of the cavity photon is $E_{c} (\kv) = \frac{\hbar c}{n_c} \sqrt{\kv^2_{\perp, c} + \kv^2} \sim \hbar \omega_0 + \frac{\hbar^2 \kv^2}{2 m_{c}}$ where $\kv_{\perp, c}$ denotes the orthogonal part of the $3$d wavevector and using the notation $\omega_0 = 2 \pi c N/(nL)$, the effective cavity photon mass $m_{c} = 2 \pi \hbar n N/(cL)$ and the exciton energy $E_{x} (p) = \varepsilon + p^2/(2 m_{x})$, which can be assumed as constant close the centre of the polariton dispersion. For our investigation, we set $E_x = 1.557$ eV, in accordance with recent results presented in \cite{Ohadi}.  $n$ is the refraction index between the cavity mirrors, $c$ the speed of light, $L$ is the cavity spacer length and $N$ the number of the quantized $z$-mode orthogonal to the $\kv$-plane.
The polariton mass of each branch depends on the effective exciton mass $m_{ x} \sim 0.1-1 m_e$ and the mass of the cavity photon $m_{c} \sim  10^{-4} m_e$ with the electron mass $m_e$ and is proportional to the inverse of the second derivative of the dispersion \eqref{bla}.

\begin{figure}[ht]
\vspace{-12mm}

\begin{tabular}{cc}
\vspace{20mm}
\begin{picture}(150,-55)
\put(-15,-180) {\includegraphics[scale=0.24]{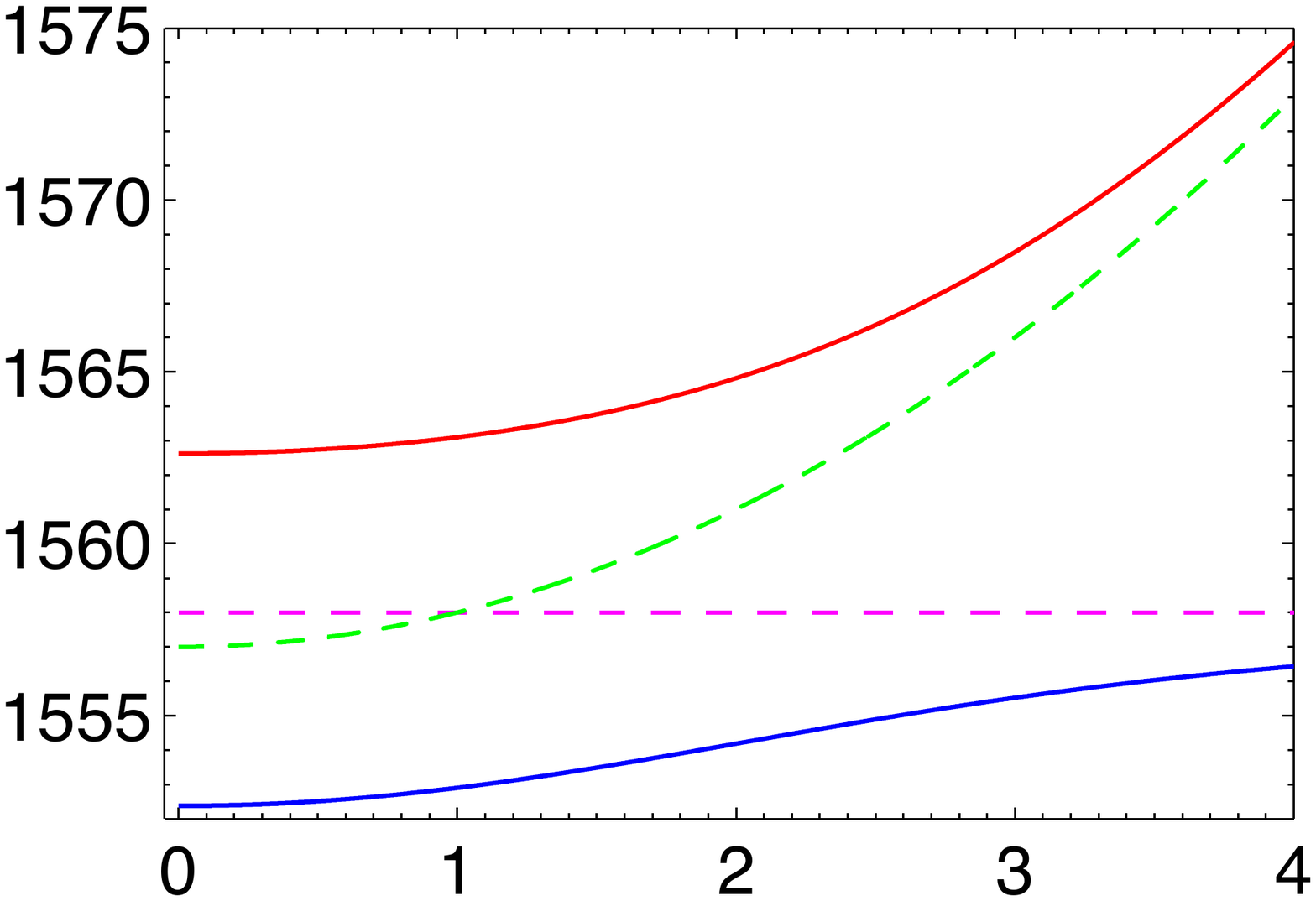} }
\put(20,-50){ \textcolor{black}{(a)}}
\put(-20,-30){ \textcolor{black}{$E (meV)$}}
\put(30,-135){ \textcolor{black}{$k  \hspace{1mm} (\mu m^{-1})$}}
\end{picture} \hspace{2mm} &
\begin{picture}(150,-80)
\put(-47,-180) {\includegraphics[scale=0.24]{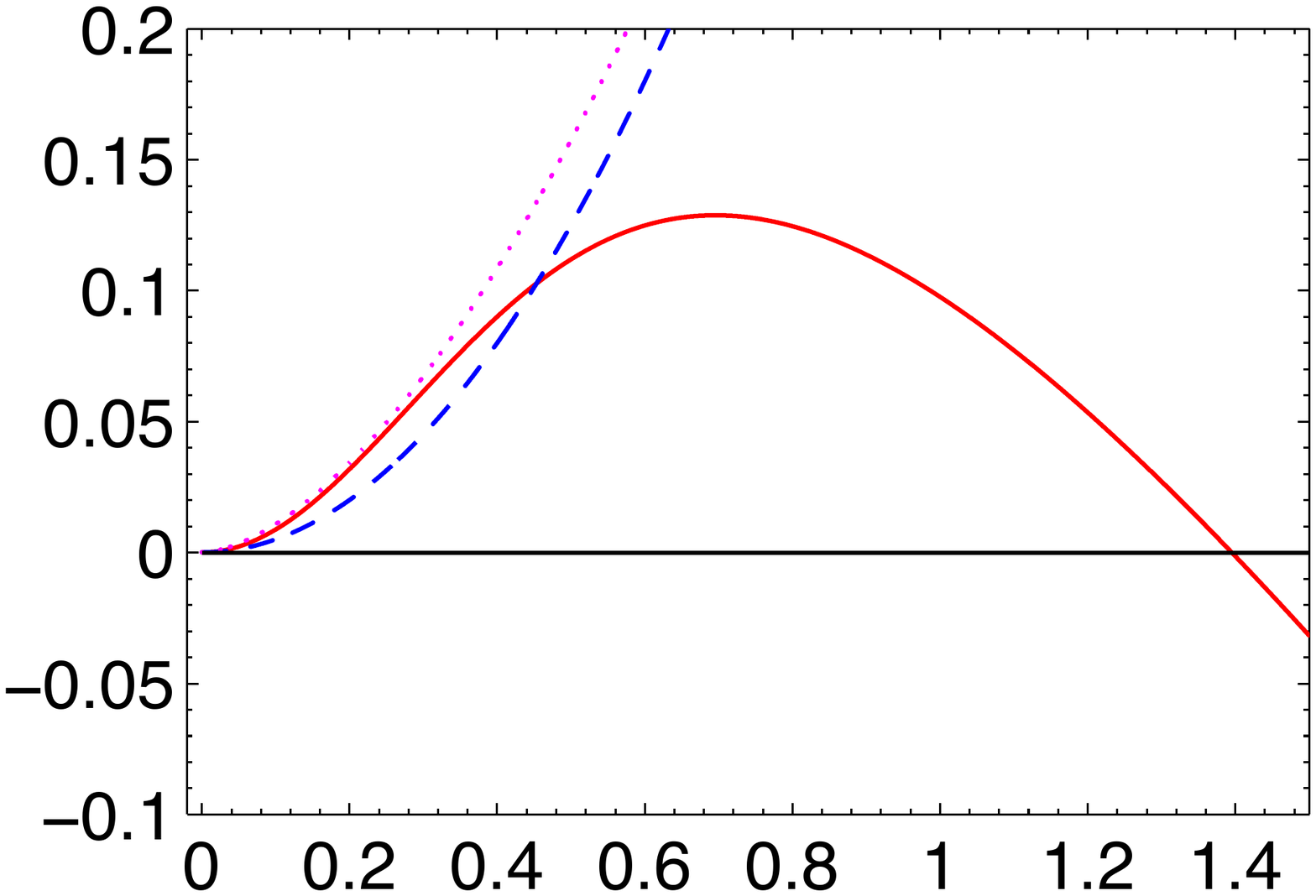} }
\put(60,-50){ \textcolor{black}{(b)}}
\put(-50,-30){ \textcolor{black}{$g (meV)$}}
\put(20,-135){ \textcolor{black}{$k  \hspace{1mm} (\mu m^{-1})$}}
\end{picture}
\end{tabular}

\vspace{25mm}

\caption{(a) Energy dispersions of the upper and lower polariton energy branch (solid lines correspondingly), the cavity (parabolic) and exciton (almost constant) dispersions (dashed lines). (b) Kinetic energy prefactors $g(k)$ defined in Eq. \ref{bla} including velocity dependent mass (solid line) compared with constant mass (dashed line) and fractional kinetic energy $k^{5/3}$ (dotted line). The sign of the kinetic energy switches for velocity dependent mass.}
\label{Num1}
\end{figure}

In Fig. \ref{Num1} (a) we show the two dispersions of the polariton, the exciton and cavity photon and in (b) the kinetic energy (prefactors) for constant, fractional and velocity dependent mass. 
Fig. \ref{Num1} (b) shows that locally fractional and generalized quantum kinetic energies are present due to the varying curvature of the effective mass - an example is given for $s=5/6$, which approximates the bottom of the polariton dispersion at $k \sim 0$ to a higher accuracy than the parabolic dispersion. The effective mass switches sign from positive at $k < k_{\rm inf}$ to negative at $k > k_{\rm inf}$. In-between it becomes infinite on a circle centered around the origin at $k =0$ in the $2$D $\kv$-plane - the inflection point $k=k_{\rm inf} \sim 1.3952 \mu m^{-1}$. While models of coupled PDEs separating the photonic and the excitonic fraction have been discussed previously \cite{Lieu,Hugo}, we present now a unifying approach for the mean-field of condensed polaritons resulting in a single non-local PDE as realistic model of generalized fractional quantum mechanics in a highly controllable solid state system.


\hspace{20mm}

\begin{figure}[ht]
\begin{tabular}{cc}
\vspace{20mm}
\begin{picture}(0,0)
\put(-130,-140) {\includegraphics[scale=0.25]{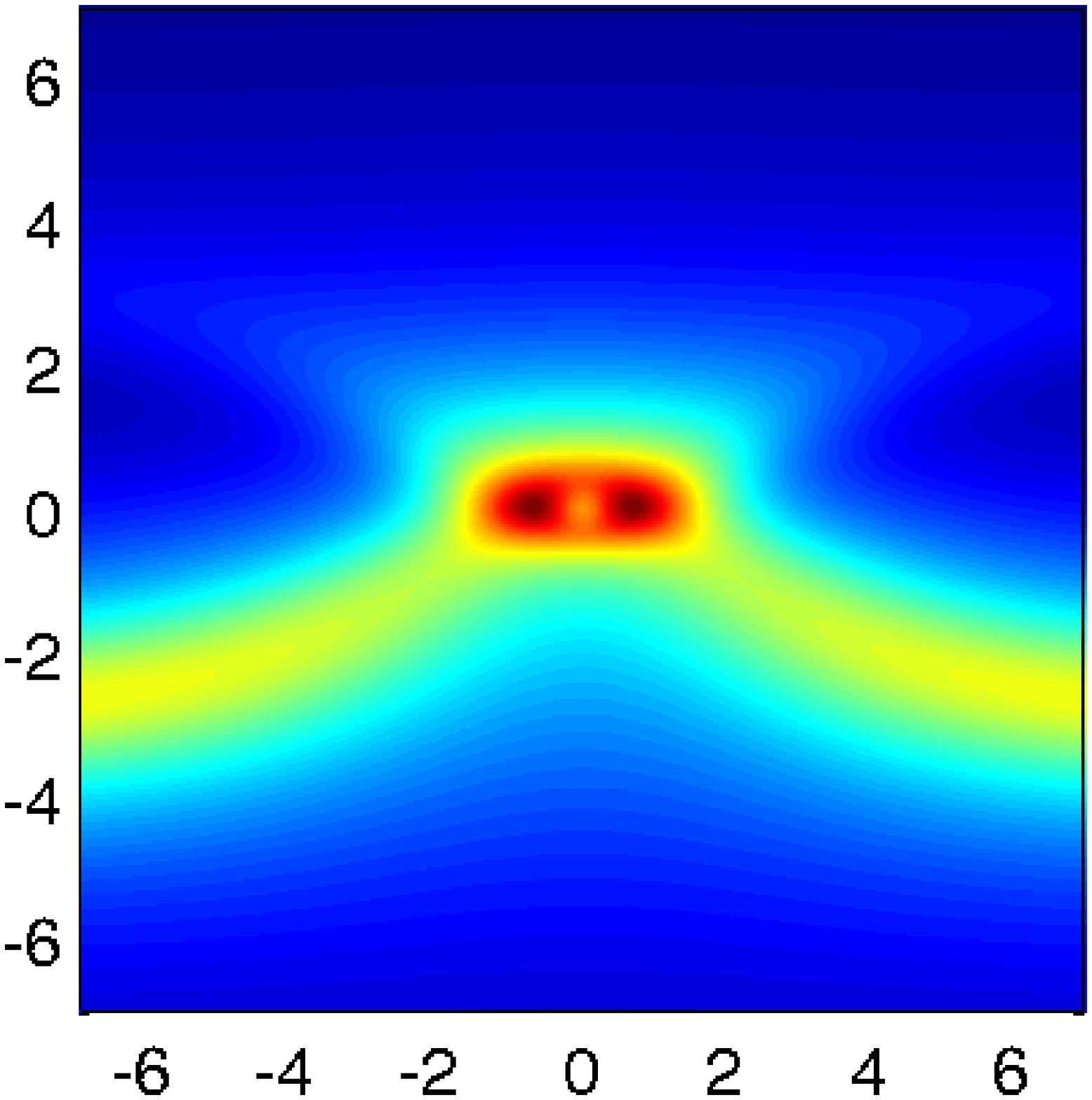} }
\put(-25,10){ \textcolor{white}{(a)}}
\put(-133,-35){ \textcolor{black}{$k_y$}}
\put(-60,-107){ \textcolor{black}{$k_x$}}
\end{picture} \hspace{2mm} &
\begin{picture}(0,0)
\put(-15,-140) {\includegraphics[scale=0.25]{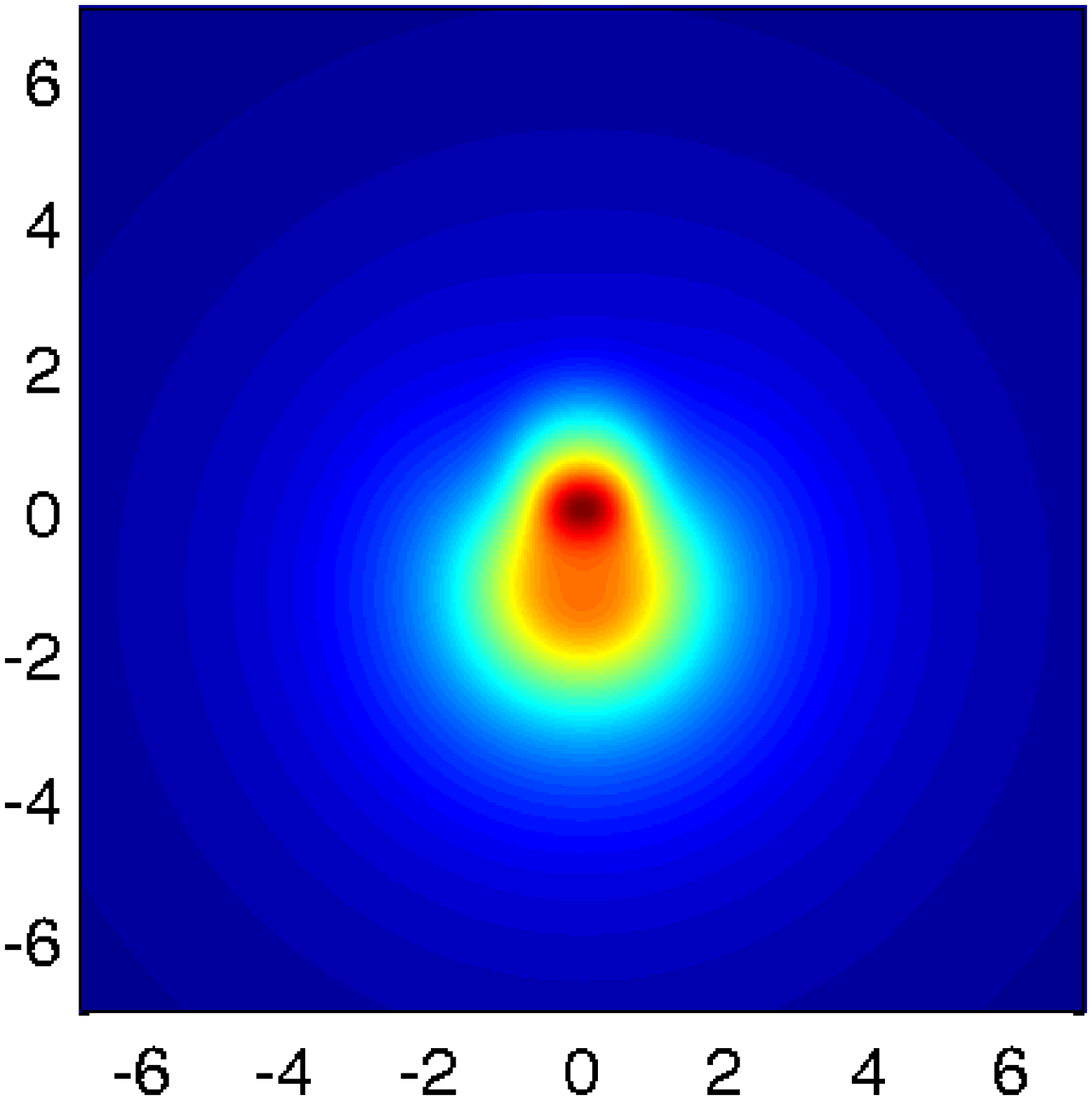} }
\put(95,10){ \textcolor{white}{(b)}}
\put(57,-107){ \textcolor{black}{$k_x$}}
\put(-15,-35){ \textcolor{black}{$k_y$}}
\end{picture}
\end{tabular}

\vspace{15mm}


\caption{(Pseudocolor plot) (a)  $|\delta \psi_{-k}|$ including velocity dependence of the effective mass for parameters $v_x =0$ and $v_y=1$ in $\kv$-space. (b) $|\delta \psi_{-k}|$ for the same parameters but for constant mass. From blue - zero magnitude to red stronger magnitude.}
\label{lin1}
\end{figure}

\emph{Generalized state equation.--}  Phenomenologically the condensate wave function is governed by a complex Ginzburg-Landau-type partial differential equation \cite{alex, Wout, jona,Pin6,Light,Rev1}, which includes the effects of polariton self-interactions, polariton-reservoir interactions and non-equilibrium properties such as gain and decay of condensate polaritons. An accurate quantum theory of polaritons is provided in \cite{east}. While in the mean-field regime the spin of polaritons can become apparent through circular polarization of the driving light source or TM-TE splitting and even spontaneously \cite{Ohadi}, we assume the spin coherent case for the introduction of the velocity dependent mass concept. We define the velocity dependent mass of the lower branch \eqref{bla} (see \cite{Dresselhaus}) and include it mathematically setting $q(\rv)=\mathcal F^{-1} [g(k)]$. Thus the kinetic energy becomes
$ \mathcal F^{-1} \left( \frac{\kv^2}{m_L (\kv)} \cdot \hat f \right) =  \mathcal F^{-1} \left( \mathcal F (q) \cdot \mathcal F (f) \right) = q \star f$ up to a constant and with $\star$ denoting a convolution.
Consequently the new polariton state equation resembling a coherent driving scheme \cite{wuc, Light} reads as follows:
\begin{multline}\label{model}
i \partial_t \psi (\rv,t)=(1-i \eta)\cdot \hbar \cdot q \star  \psi (\rv,t)+i P (\rv,t) -i \frac{\gamma}{2} \psi(\rv,t)   \\ + \left[ \alpha |\psi|^2 + V(\rv ,t)  + \omega \right] \psi (\rv,t)
\end{multline}
The frequency of the bottom of the lower brach is denoted $\omega$, $\alpha$ is the self-interaction strength, $V(\rv,t)$ is an external potential, while $\gamma$ is the loss rate of polaritons due to their decay. The coherent pumping field is \cite{wuc,Light}
\beq
P(\rv,t) = P_0(\rv) e^{i \kv_{\rm i}\cdot\rv} e^{-i \omega_{\rm i} t}
\eeq
with $P_0$ denoting the pump profile amplitude, $\kv_{\rm i}$ denoting the $2$d pump wave vector and $\omega_{\rm i}$ the pump frequency. Fig. \ref{Num1} (b) shows that locally close to the particular $\kv$ under consideration fractional kinetic energies emerge and thus the presented model \eqref{model} includes as a special case a feasible implementation of (driven) fractional quantum mechanics.
Via $\kv_{\rm i}$ we can choose experimentally at which particular $\kv$ the condensate wave function is formed on the dispersion in $\kv$-space. Alternatively with incoherent driving schemes one could control $\kv$ of the condensate by adjusting the spot size, which determines the final velocity of condensed polaritons  \cite{opt}.

\emph{Plane waves of the velocity dependent  model.--} Let us now present analytical plane wave expressions respecting the velocity dependent mass of the polariton system.  We consider a $1$d scenario, assume homogeneous pumping, no external potential $V=0$ and set as ansatz for the stationary solution $\psi (x,t) = \psi_0 e^{i \kappa_{\rm i} x} e^{-i \omega_{\rm i} t}$,
\begin{eqnarray}
&&\hbar  \cdot ( q \star \psi_0 e^{i \kappa_{\rm i} x} )
 e^{-i \kappa_{\rm i} x} + \left[ \omega_{\rm i}   - \alpha |\psi_0|^2
 - \omega + i \frac{\gamma}{2} \right] \psi_0 (x)\nonumber\\
&&\quad =  i P_0.
\end{eqnarray}
As discussed in the supplemental material \cite{sup} by setting $\hat \psi (k - \kappa_{\rm i})
= \psi_0 \delta (k - \kappa_{\rm i})$ we obtain three analytical solutions \cite{sup}. The physical solution is
\begin{multline}
\psi^{+}_0 = \\ =    \frac{(1+ i \sqrt{3}) (a+ i b)}{ 2^{2/3} (-27 i \alpha^2 P_0 + \sqrt{4(-3a \alpha-3ib \alpha)^3 -729 \alpha^4 P^2_0})^{1/3}} - \\  \frac{(1- i \sqrt{3}) (-27 i \alpha^2 P_0 + \sqrt{4 (- 3 a \alpha - 3 i b \alpha)^3 - 729 \alpha^4 P^2_0})^{1/3}}{ 6 \cdot 2^{1/3}}
\end{multline}
using the abbreviations $a = \omega_{\rm i} - \omega  -   \frac{ \kappa_{\rm i}^2}{m_L ( \kappa_{\rm i})} \hbar$ and $b = \gamma/2$. The density $\rho^+ = |\psi^+_0|$ tends to zero for $P_0 \to 0$ corresponding to no pumping of polaritons into the condensate and it increases monotonically with $P_0$.  The solutions are modified by the velocity dependent mass  \eqref{bla} and the constant mass case is obtained by substituting $m_L (\kv) \to m_L$. For the plane wave scenario respecting $m(k)$ implies including its value at $\kappa_{\rm i}$ of the dispersion (see Fig. \ref{Num1} (b)), which modifies the magnitude of the wave function (or the luminosity of the micro-cavity). In particularly the solutions respect the negative mass effect and note that the $1$d plane wave solutions can be trivially extended to $2$d.  Nonlocal effects can be expected in more general pumping schemes as shown below. 

\emph{Linear waves analysis for the generalized driven system.--} The linearized elementary excitation equation around the stationary states without trapping and under coherent pump with $k_{\rm i} =0$, for $\psi(\rv,t) = \sqrt{\rho} \exp(-i \omega_{\rm i} t)$ is stated in \cite{Wout, Light}
and the eigenvalues of the Bogoliubov operators for the plane wave modes respecting the $m(k)$ dependence are the generalized Bologliubov dispersions of excitations as stated in the supplemental material \cite{sup}.
The signs of the dispersion correspond to the positive and negative Bogoliubov branch and as expected the presence of the velocity dependent mass adapts the excitation energy by a non-parabolic $k$ dependence of the effective kinetic energy.

 The driven Gross-Pitaevskii equation \eqref{model} without external potential, energy relaxation (with effects discussed mathematically in \cite{Pini}), absorbing $\omega$ into the phase and setting $\hbar =1$ reads
\begin{equation}\label{model2}
 i \frac{\partial \psi}{\partial t} = q \star
 \psi  + \alpha |\psi|^2 \psi - i \gamma \psi + i P.
\end{equation}
To consider the movement of linear waves with velocity $\vv$ we make an ansatz of the form $\psi = \phi_0 + \delta \psi$, where $\phi_0$ represents the unperturbed part solving the generalized driven GPE and $\dpsi$ a small perturbation on top of it. We can think of $\phi_0$ as one of the solutions derived in the previous section. By inserting this ansatz in $\eqref{model2}$, introducing a chemical potential $\mu$ (to be identified later) and dropping terms of order $\dpsi^2$ we get the Bogoliubov equation for the perturbation as discussed in more detail in the supplemental material \cite{sup}.  We denote the Fourier transform of the pump as $\tilde P = i \int e^{-i \kv\cdot \rv } (P - \gamma \phi_0)$  set w.l.o.g.  $\phi_0 = \sqrt{n} \exp(i \theta_0)$ and choose the constant phase of the pump such that $\phi_0 \to \sqrt{n}$  and define  $\mu = \alpha n$. Thus  in the reference frame moving with the linear wave with velocity $\vv$ on top of the condensate $\phi_0$ we get the explicit solution
\begin{equation}\label{solution1}
\dpsi^*_{-\kv} =  \frac{ \tilde P \mu + \tilde P^* ( i \gamma+  \kv\cdot
\vv - g(|\kv|) -\mu )}{\gamma^2+ 2 \mu g(|\kv|)  + g(|\kv|)^2 - 2 i \gamma \kv\cdot \vv - (\kv\cdot \vv)^2}.
\end{equation}
This solution is a natural extension of the equilibrium atomic BEC solutions presented in \cite{pita,glad} and those for constant mass discussed in \cite{Light, Hugo}.
In Fig. \ref{lin1} we compare the solutions $\dpsi^*_{-\kv}$ due to constant and velocity dependent mass. For the sake of simplicity we assume $\tilde P =1$, $\gamma =1$, $\mu=1$ and $g(k)$ is given as described above. The results indicate a significant difference in the linear wave condensate dynamics which particularly will be investigated numerically in more detail in the following section.

\begin{figure}[ht]
\vspace{-30mm}
\begin{tabular}{cc}
\vspace{35mm}
\begin{picture}(200,0)
 \put(-29,-150){\includegraphics[width=150pt, height=200pt]{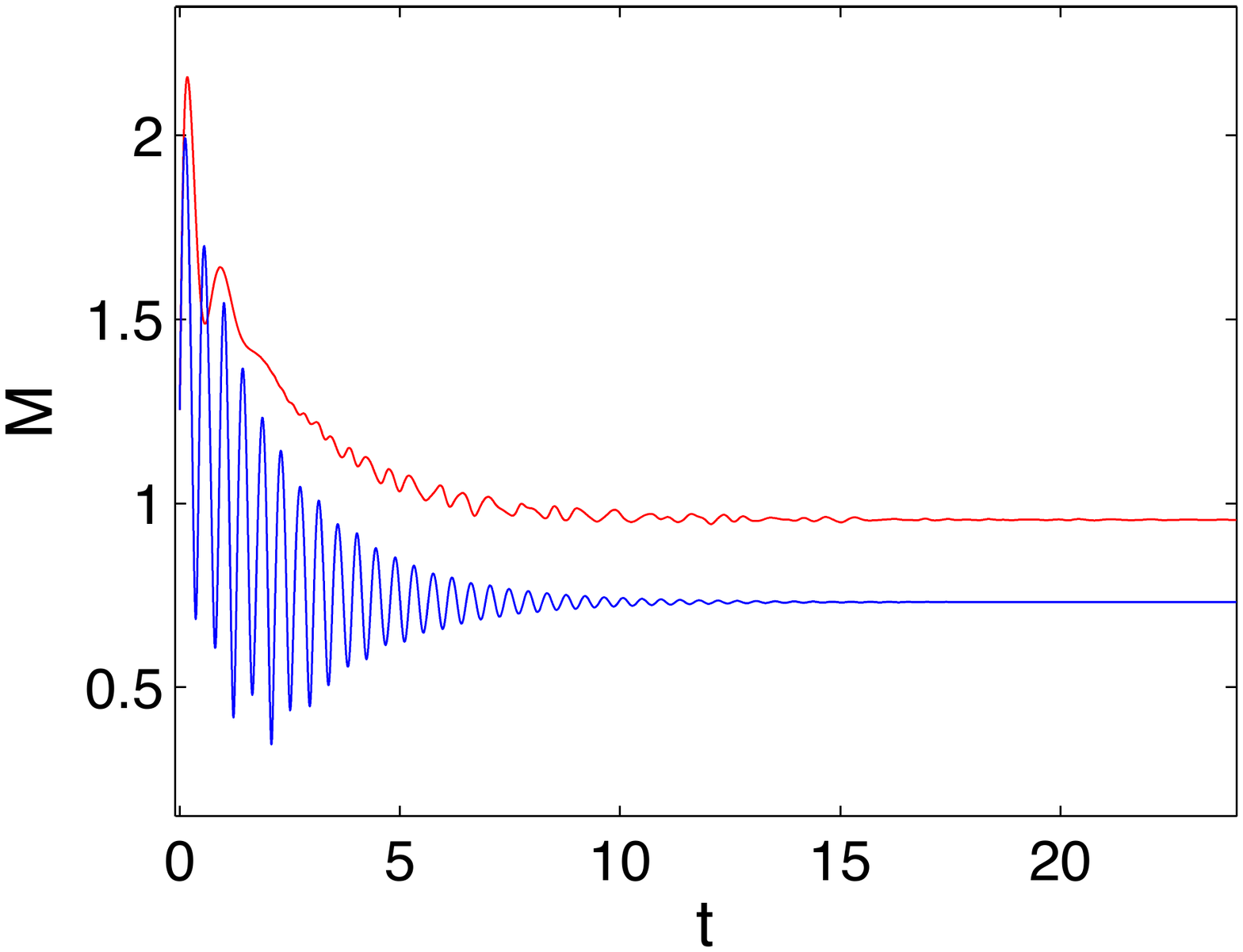} }
\put(87,-18){ \textcolor{black}{(a)}}
\vspace{10mm}
\end{picture} &
\begin{picture}(200,100)
 \put(-96,-150){\includegraphics[width=139pt, height=190pt]{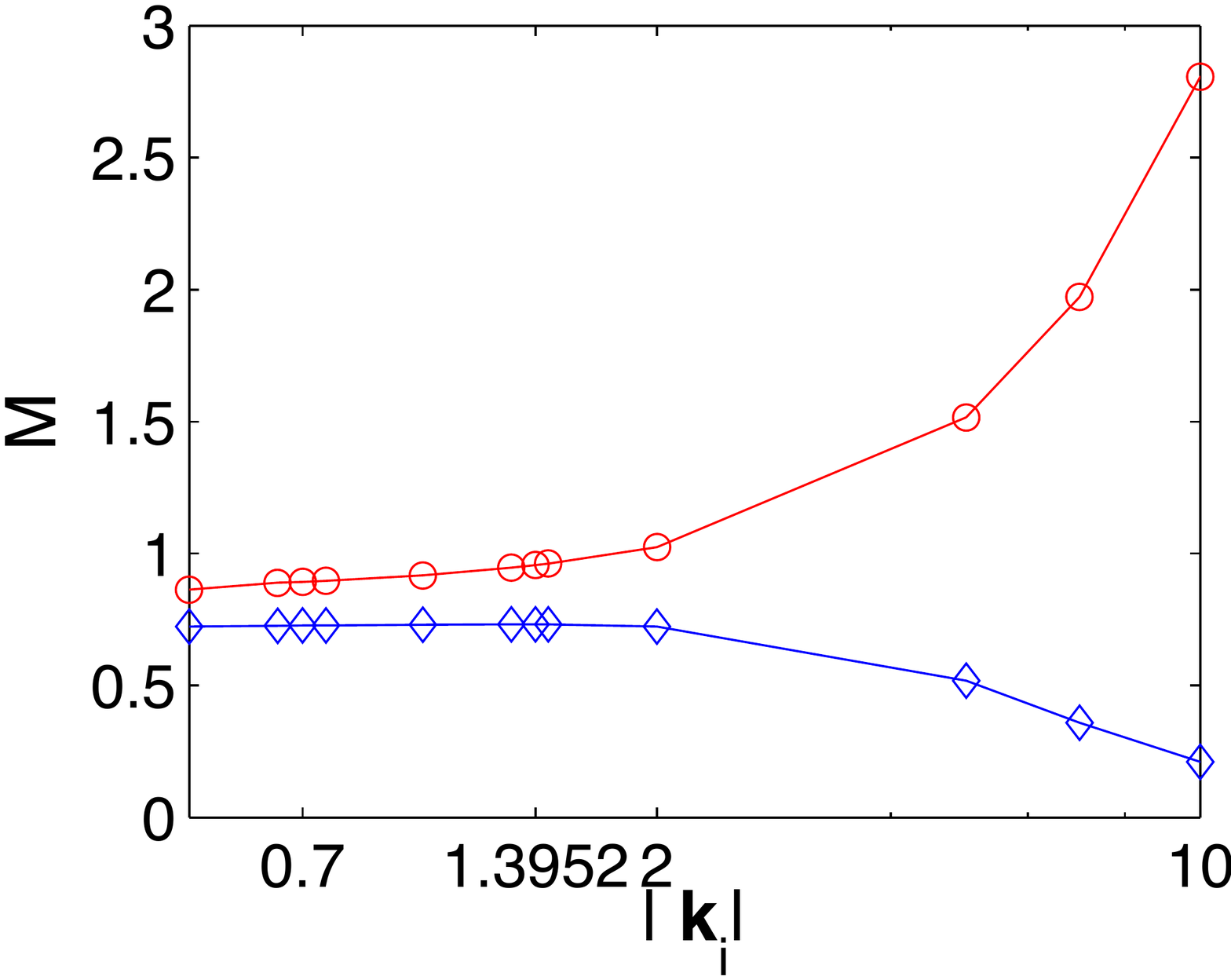} }
 \put(20,-18){ \textcolor{black}{(b)}}
\end{picture}
\end{tabular}
\vspace{0mm}
\caption{(a) Red line corresponds to the $M$ including the $m(k)$ effect over time $t$ with $[t] =  ps$. The blue line corresponds to the model with constant mass. Parameters are outlined in \cite{Fig3}.  \label{Fig3} (b) The red line corresponds to $M$ including the $m(k)$ effect. The blue line corresponds to the model with constant mass \cite{Fig4}. Units are $[\kv_{\rm i}] = \mu m^{-1}$ while the $M$ corresponds to the total occupation number.   \label{Fig4}}
\end{figure}



\begin{figure}[ht]
\begin{tabular}{c}
\begin{picture}(200,100)
 \put(-28,70){\includegraphics[width=250pt, height=125pt]{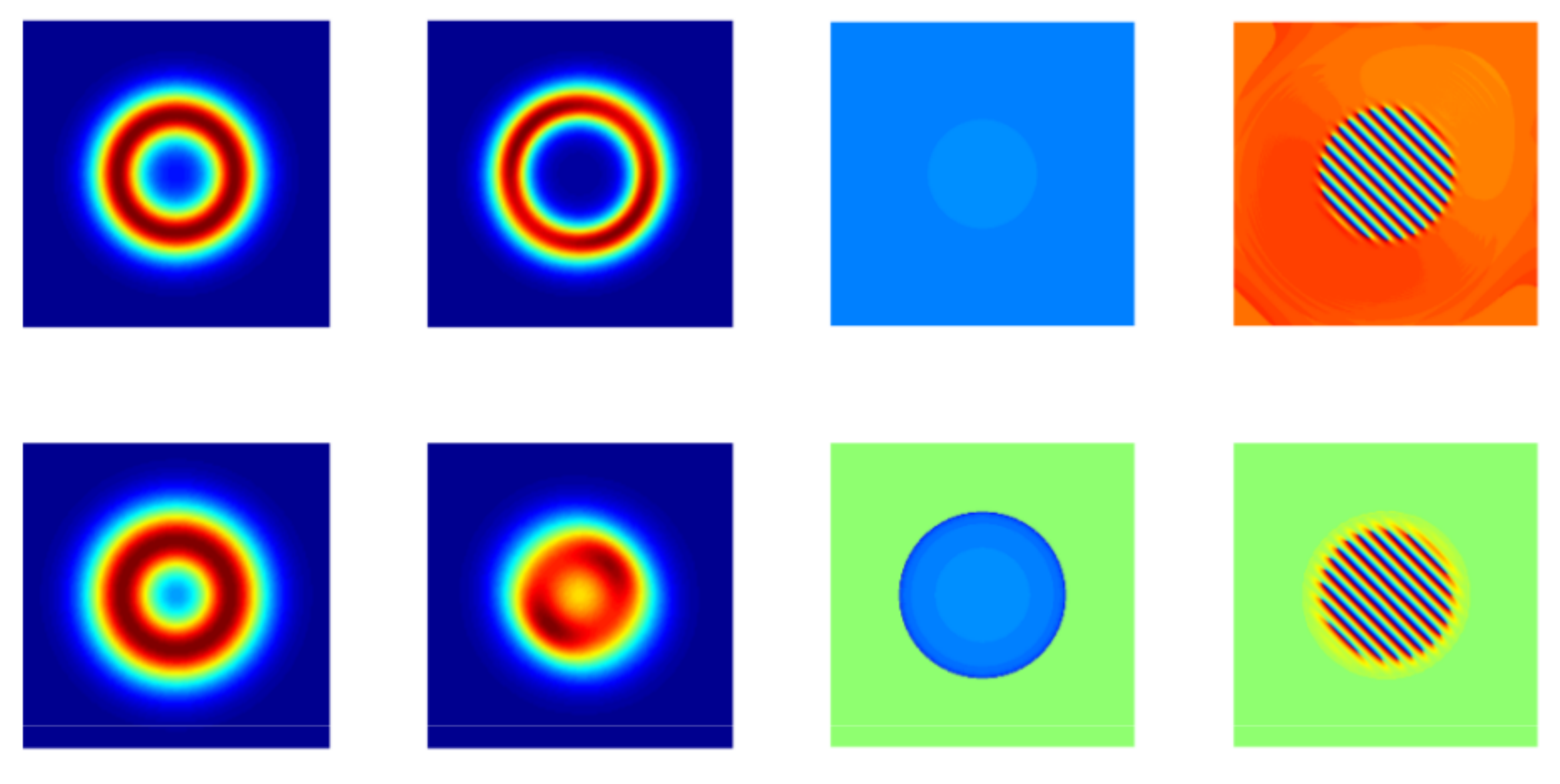} }
  \put(8,180){ \textcolor{white}{(a)}}
    \put(72,180){ \textcolor{white}{(b)}}
      \put(130,180){ \textcolor{white}{(aa)}}
    \put(195,180){ \textcolor{white}{(bb)}}
     \put(8,113){ \textcolor{white}{(c)}}
    \put(72,113){ \textcolor{white}{(d)}}
          \put(130,113){ \textcolor{black}{(cc)}}
    \put(195,113){ \textcolor{black}{(dd)}}
\end{picture}
\end{tabular}
\vspace{-25mm}
\caption{(Pseudocolor plots) Comparison between the $2$D wave functions governed by the $m(k)$ respecting model in the upper line - (a) density and (aa) phase for $a=0$ and (b) density and (bb) phase for $a=10.38$. Correspondingly in the lower line, density (c), (d) and phase (cc), (dd) stemming from the classic model.  Blue corresponds to lower densities and red is associated with higher densities. Parameters chosen as in \cite{Fig5}. \label{Fig5}}
\end{figure}
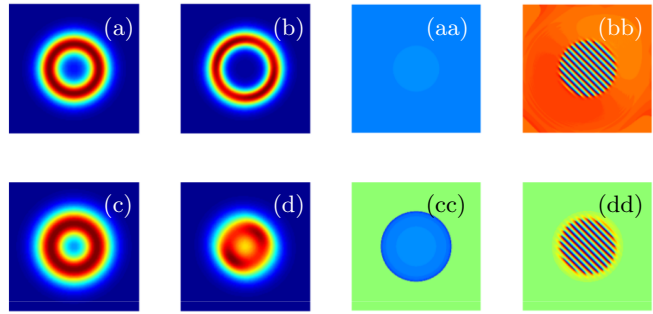

\emph{Numerical experiments.--} We consider  for Eq. $\ref{model}$ the $2$d scenario of harmonic trapping $V (r) = k r^2$ with strength $k$ and a Gaussian pump profile $P_0(x) = A \exp{-\frac{(|\rv-{\bf d}|)^2}{\sigma^2}}$, where $A$ is the amplitude, ${\bf d}$ denotes the position of the pump and $\sigma$ its width. In Fig. \ref{Fig3} (a) we show a comparison of the total mass of a perturbed condensate wave function $\| \psi \|^2_2 = \int_{\mathbb R^2} |\psi|^2 d \rv = M$ ($L^2$-norm) as it varies in time. 
To generate the perturbation we start with an initial condition $\psi_I (x,y) = \exp(-x^2+y^2)$ and evolve it according to \eqref{model} for $m=const.$ and $m = m(k)$. While constant mass implies oscillations of the $L^2$-norm/total density at a given instance of time the velocity dependence of the mass acts as a damping term as shown in Fig. \ref{Fig3} (a). For larger times both models induce convergence to a stationary state of different $L^2$-norm. To elucidate the differences in the total mass of the condensate wave function we show in Fig. \ref{Fig4} (b) a comparison for different $\kv_{\rm i} (\mu m^{-1})$. As  $\kv_{\rm i}$ increases $M$ monotonically increases for the $m(k)$ model while it decreases for the classic model, hence offering an experimentally feasible test of the new theory \eqref{model}. Furthermore in Fig. \ref{Fig5} we show a comparison of the $2$d density distributions of stationary states for different $\kv_{\rm i} = \sqrt{2}/2(a,a)$ with $a \in\{0,10.38 \}$, $[\kv] = \mu m^{-1}$ and accordingly chosen $ \omega_{\rm i}$ in a mexican hat potential (i.e. harmonic trapping with a tight gaussian at the center \cite{Fig5}). We observe that the density cloud/luminosity for velocity dependent mass contracts as $k$ increases (see as well \cite{sup}) - a behavior analog to attractive atomic BEC in traps  \cite{revBEC}. Instead of negative/attractive self-interactions, the negative mass induces a relative sign between the kinetic energy and the still repulsive self-interactions \eqref{model} leading to the observed contraction consistent with the experimental results in \cite{Schef}, The diameter of the ring shaped condensate increases with $k$ for the $m(k)$ model Fig. \ref{Fig5} (a), (b) while it decreases for the classic theory (c), (d). In addition we simulate \eqref{model} reduced to $1$d and obtain the qualitative results presented \cite{sup} again showing the increase of total mass. Furthermore we provide in \cite{sup} more detailed numerical results for the case with and without harmonic trapping.
\vspace{35mm}

\emph{Conclusions.--}  We have identified the polariton condensate wave functions with those of fractional quantum mechanics by respecting the velocity dependent mass of polaritons in the governing PDE. More generally because the $\kv$ dependent kinetic energy of the polariton condensate deviates significantly from the parabolic form new phenomena could be observed. Remarkably for $k \sim 0$ a fractional nonlinear Schr\"odinger-type equation is the more accurate model compared to the classic parabolic nonlinear SE type models previously used. Via a feasible coherent pumping scheme - driving the polariton to condense at a chosen single point of its dispersion - one can effectively switch between different points of the energy dispersion enabling to test the effects of velocity dependent mass. The importance of the in-plane momentum for the emerging polariton condensate was shown in the explicit analytical expressions stated and further the linear waves analysis suggests significant different dynamical behaviour. Consequently numerical simulations for $1$d and $2$d scenarios for which a novel time splitting Fourier pseudospectral method has been developed revealed evident differences in the predictions for the polariton condensate to the classic results. The dynamics show a suppression of total density oscillations due to the velocity dependence of the mass, the total mass increases for larger $k$ while classic mean field models predict a loss. The latter is a phenomenon is a feasible test of the theory presented here. In addition condensates forming above the inflection point show attractive-type density profiles in accordance with the observations in \cite{Schef}. While the coherent driving scheme utilized in this paper pre-defines the phase and suppresses the spontaneous emergence of excitations such as vortices or dark and bright solitons incoherent driving schemes may reveal interesting pattern formation in future works.

\emph{Acknowledgements.--} F.P. acknowledges financial support by the UK
Engineering and Physical Sciences Research Council (EPSRC) through his
doctoral prize fellowship at the University of Cambridge and his
Schr\"odinger Fellowship at the University of Oxford.
W.B. acknowledges support by
the Ministry of Education of Singapore grant R-146-000-196-112,
and Y.Z. acknowledges support by
the Natural Science Foundation of China grant 91430103 and 11471050.
We are very grateful to Peter Cristofolini, Jesus Sierra and Peter
Markowich for stimulating discussions and a keen interest in this topic.


\begin{thebibliography}{10}
\bibitem{Fig4} To generate the results we used the same model and numerical parameters as \cite{Fig3}, except a varying pump vector $\kv_{\rm i} = \sqrt{2}/2(a,a)$.
\bibitem{Fig5} To generate the results we used the same model and numerical parameters as \cite{Fig4}, but added to the harmonic trapping a tight gaussian with the same origin, i.e. $V(\rv) = |\rv|^2 + 50 \exp(- |\rv|^2/0.2)$.
\end{thebibliography}
 \end{document}